\documentclass[doublecol]{epl2} 

\usepackage{hyperref}
\usepackage{amsmath}
\usepackage{amssymb}

\title{When does cyclic dominance lead to stable spiral waves?}

\author{
	Bartosz Szczesny\inst{1}\thanks{Electronic address: \email{mmbs@leeds.ac.uk}} \and
	Mauro Mobilia\inst{1}\thanks{Electronic address: \email{M.Mobilia@leeds.ac.uk}} \and
	Alastair M. Rucklidge\inst{1}\thanks{Electronic address: \email{A.M.Rucklidge@leeds.ac.uk}}
}
\shortauthor{Szczesny, B. \etal}

\institute{
	\inst{1} Department of Applied Mathematics, School of Mathematics, University of Leeds, Leeds LS2 9JT, U.K.
}

\abstract{
Species diversity in ecosystems is often accompanied by the self-organisation of the population into fascinating spatio-temporal patterns. Here, we consider a two-dimensional three-species population model and study the spiralling patterns arising from the combined effects of generic cyclic dominance, mutation, pair-exchange and hopping of the individuals. The dynamics is characterised by nonlinear mobility and a Hopf bifurcation around which the system's phase diagram is inferred from the underlying complex Ginzburg--Landau equation derived using a perturbative multiscale expansion. While the dynamics is generally characterised by spiralling patterns, we show that spiral waves are stable in only one of the four phases. Furthermore, we characterise a phase where nonlinearity leads to the annihilation of spirals and to the spatially uniform dominance of each species in turn. Away from the Hopf bifurcation, when the coexistence fixed point is unstable, the spiralling patterns are also affected by nonlinear diffusion.
}

\begin{document}

\maketitle

\section{Introduction}
In nature, organisms live in areas much larger than the distances they typically travel and thus they interact with a finite number of individuals in their neighbourhood. Space and mobility are therefore crucial ingredients in understanding how populations evolve and how ecosystems self-organise. Even in the presence of sources of randomness and inhomogeneities, spatial degrees of freedom and movement can lead to the formation of characteristic spatio-temporal patterns~\cite{spatio-temporal}, whose origin in ecosystems has been a subject of intense research for decades~\cite{spatio-temporal,patterns,PatternsRPS}. In his pioneering work, Turing showed that pattern-forming instabilities can be caused by diffusion~\cite{Turing}. While Turing patterns have been found in ecology and biology~\cite{patterns}, the requirements of Turing's theory (e.g. separation of scales in diffusivities) appear to be too restrictive to explain pattern formation in many ecosystems, see e.g. Ref~\cite{TuringNoisy}.

Another important problem concerns the mechanisms promoting the maintenance of biodiversity~\cite{biodiversity}. In this context, cyclic dominance has been recently proposed as an intriguing motif facilitating the coexistence of diverse species in ecosystems. Examples of cyclic competition between three species can be found in coral reef invertebrates, {\it Uta stansburiana} lizards, and communities of {\it E.coli}~\cite{PatternsRPS,Kerr,Nahum,LizardMutation}. In the experiments of Ref.~\cite{Kerr}, the cyclic competition of three bacterial strains on two-dimensional plates was shown to yield patterns sustaining species coexistence. Such competition is metaphorically described by {\it rock-paper-scissors} (RPS) games, where ``rock crushes scissors, scissors cut paper, and paper wraps rock''~\cite{games}. While non-spatial RPS-like models often evolve towards extinction of all but one species in finite time~\cite{RPSnonspatial}, their spatial counterparts are generally characterised by the long-term coexistence of species and by the formation of complex spatio-temporal patterns~\cite{zerosum,RMF,RF,Matti,He2}. Recently, various two-dimensional versions of the model introduced by May and Leonard~\cite{MayLeonard} have received much attention~\cite{RMF,RF,Matti,He2}. When mobility is implemented by pair-exchange among neighbours, species coexistence is long-lived and populations form non-Turing spiralling patterns below a certain mobility threshold, whereas biodiversity is lost when that threshold is exceeded~\cite{RMF}.

In this Letter, we characterise the intricate patterns emerging from the dynamics of a {\it generic} model of a cyclically competing three-species population, and study how these patterns affect the maintenance of biodiversity in two dimensions. The basic evolutionary processes considered here are the most general form of cyclic dominance between three species obtained by combining and unifying the interactions of Refs.~\cite{RMF,RF,He2,Jonas}. Inspired by the experiments of~\cite{Nahum}, the model is formulated at the metapopulation level~\cite{metapopulation}, and is characterised by a Hopf bifurcation as well as by a form of movement that discriminates between crowded and dilute regions and results in nonlinear mobility. While spiralling patterns have often been observed numerically in related models~\cite{RMF,RF,Matti,He2}, we here demonstrate that nonlinearity and mobility can disrupt the stability of the ensuing spiral waves. Our main result is the phase diagram derived from a {\it controlled} perturbative multiscale expansion around the Hopf bifurcation. The diagram is characterised by three phases (parameter regimes) in which spiral waves are unstable, and by one phase where spiralling patterns are stable. In one of these unstable phases, spiral waves annihilate and each species dominates the system in turn.

\section{Model}
The generic model of cyclic dominance between three competing species is defined on a periodic square lattice of $L^2$ patches, $L$ being the linear size, labelled by a vector ${\bm l}=(l_1,l_2)$~\cite{FigshareMovies}. Each patch has a limited carrying capacity, accommodating at most $N$ individuals, and consists of a well-mixed population of species $S_1, S_2, S_3$ and empty spaces $\O$. Within each patch ${\bm l}$, the population composition evolves according to
	\begin{eqnarray}
	\label{sel}
		S_i + S_{i+1} \xrightarrow{\sigma} S_i + \O &~&
		S_i + S_{i+1} \xrightarrow{\zeta } 2 S_i \\
	\label{non-sel}
		S_i + \O \xrightarrow{\beta} 2 S_i &~&
		S_i \xrightarrow{\mu} S_{i\pm1},
	\end{eqnarray}
where the species index $i \in \{1,2,3\}$ is \textit{ordered cyclically} such that $S_{3+1} \equiv S_1$ and $S_{1-1} \equiv S_3$. The reactions (\ref{sel}) describe the cyclic competition between the species: $S_i$ dominates over $S_{i+1}$ while being dominated by $S_{i-1}$. Here, we consider a generic form of cyclic competition by separating the zero-sum process of dominance-replacement (rate $\zeta$), as studied in Ref.~\cite{RF}, from the dominance-removal selection process (rate $\sigma$) of Refs.~\cite{RMF, He2}. With reactions (\ref{non-sel}), we assume that births (rate $\beta$) occur independently of the cyclic competition provided that space is available~\cite{Nahum}. It should be noted that without loss of generality any of the rates of (\ref{sel},\ref{non-sel}) can be set to one by properly defining the time scale. To illustrate our results we shall here make the choice to set $\beta=1$.

In addition, we also assume that each species can mutate into one another (rate $\mu$). Such mutations have been found in some of the ecosystems that have inspired our model. For example, the {\it E.coli} bacteria are known to mutate \cite{Kerr} and the side-blotched lizards {\it Uta stansburiana} have been found to undergo throat-colour transformations~\cite{LizardMutation}. Below, we show that a non-zero mutation rate ensures that the model exhibits a Hopf bifurcation, which is a feature on which our analysis builds.

As biological movement is often nonlinear and driven by local population density~\cite{Kearns}, we here {\it divorce} hopping (rate $\delta_D$) from pair-exchanges (rate $\delta_E$) between nearest-neighbour patches ${\bm l}$ and ${\bm l'}$~\cite{He2}, according to
	\begin{eqnarray}
	\label{migr}
		\big[S_i \big]_{\bm l} \big[\O \big]_{\bm l'} &\xrightarrow{\delta_D}&
		\big[\O	\big]_{\bm l} \big[S_i \big]_{\bm l'} \nonumber \\
		\big[S_i \big]_{\bm l} \big[S_{i\pm1} \big]_{\bm l'} &\xrightarrow{\delta_E}&
		\big[S_{i\pm1} \big]_{\bm l} \big[S_i \big]_{\bm l'},
	\end{eqnarray}
where ${\bm l}$ and ${\bm l'}$ lie in 4-neighbourhood. The processes (\ref{migr}) lead to nonlinear mobility (see (\ref{PDE}) below) and allow us to distinguish the movement in crowded regions, where pair-exchange dominates, from mobility in dilute systems, where hopping is more likely. The metapopulation model (\ref{sel})-(\ref{migr}) is well-suited to capture stochastic effects via size expansion in the carrying capacity and allows a natural connection with its deterministic description~\cite{TuringNoisy,inprep,SizeExp}.

It has to be noted that most previous works considered lattice models with $N=1$ and nearest-neighbour reactions (\ref{sel})-(\ref{non-sel}), while here these interactions occur on-site. Apart from these differences, the processes that we consider are similar to those of~\cite{RMF} in the special case where $\zeta=\mu=0$ and $\delta_D=\delta_E$, while some aspects of the system's properties with $\zeta \ne 0$, $\mu \ne 0$ and $\delta_D \ne \delta_E$ have been investigated in~\cite{RF},~\cite{Jonas} and~\cite{He2}, respectively.

\section{Dynamics and size expansion}
When $N \to \infty$, the leading-order term in the size expansion yields mean field rate equations for the continuous species densities $s_i= N_{S_i}/N$~\cite{inprep,SizeExp}. Here, $N_{S_i}$ is the number of $S_i$'s in one patch. With ${\bm s}\equiv (s_1, s_2, s_3)$,
	\begin{eqnarray}
		\label{RE}
		\frac{d s_i}{d t}
		&=& s_i [\beta (1-r) - \sigma s_{i-1} + \zeta (s_{i+1} - s_{i-1})] \nonumber \\ 
		&~& + \mu (s_{i-1} + s_{i+1} - 2 s_i) \equiv {\cal F}_i(\bm{s}),
	\end{eqnarray}
where $r\equiv s_1+s_2+s_3$ is the total density. 
It is worth noting that the hopping/exchange processes (\ref{migr}) do not appear in these mean field equations that ignore the spatial degrees of freedom. Eqs.~(\ref{RE}) admit a coexistence fixed point ${\bm s}^*=\frac{\beta}{3\beta + \sigma}(1,1,1)$. In the presence of mutations, ${\bm s}^*$ is an asymptotically stable focus when $\mu > \mu_H = \frac{\beta\sigma}{6(3\beta + \sigma)}$, while there is a supercritical Hopf bifurcation (HB)~\cite{Jonas} at $\mu=\mu_H$ and a stable limit cycle of frequency $\omega_H \approx \frac{\sqrt{3}\beta (\sigma + 2\zeta)}{2(3\beta + \sigma)}$ when $\mu<\mu_H$. For later convenience, the departure from the HB point is measured by a parameter $\epsilon$ defined by $\mu = \mu_H - \frac{1}{3}\epsilon^2$. In stark contrast, when $\mu=0$ (no mutations), the coexistence state ${\bm s}^*$ is never asymptotically stable. Instead, solutions of (\ref{RE}) are either heteroclinic cycles ($\mu=0$ and $\sigma>0$)~\cite{MayLeonard} or nested neutrally stable periodic orbits (in the special case $\mu=\sigma=0$)~\cite{games}. In either case, finite-size fluctuations cause the rapid extinction of two of the three species in a non-spatial setting~\cite{RPSnonspatial}.

	\begin{figure}
		\includegraphics[width=0.99\linewidth]{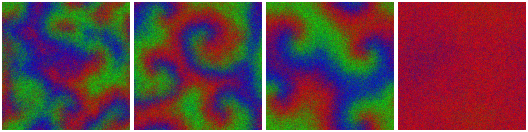}
		\caption{Reactive steady states in stochastic Gillespie simulations of reactions (\ref{sel})-(\ref{migr}). Here, $L^2 = 128^2$, $N=64$, $\beta = \sigma = \delta_D = \delta_E = 1$, $\mu = 0.02 < \mu_H = 0.042$ ($\epsilon \approx 0.26$) and, from left to right, $\zeta=(1.8, 1.2, 0.6, 0)$. Each pixel describes a patch with normalized {\tt RGB} representation $(red,green,blue)=(s_1,s_2,s_3)$ of its state. The right-most panel shows an oscillatory homogeneous state in which each of the species dominates the whole population in turn (see Fig.~\ref{ssa_spiral_annihilation} for time evolution). Initially $\bm{s} \approx \bm{s}^*$ with small random perturbations, see~\cite{FigshareMovies}.}
		\label{ssa_four_phases}
	\end{figure}

When spatial dependence is taken into account in the limit $L \to \infty$ and lattice spacing~$\to 0$, the spatial coordinate ${\bm x}\equiv \bm{l}/L$ becomes continuous. The densities thus depend on space and time, $s_i\equiv s_i({\bm x},t)$, and obey
	\begin{equation}
		\label{PDE}
		\partial_t{s}_i = {\cal F}_i(\bm{s})+\delta_D \Delta s_i
		+ (\delta_D-\delta_E) \left(s_i \Delta r - r \Delta s_i \right),
	\end{equation}
where $\Delta=\partial_{x_1}^2+\partial_{x_2}^2$ and the nonlinear diffusive terms ($s_i \Delta r -r\Delta s_i$) arise from the divorce between pair-exchange and hopping~\cite{He2,Fanelli}. With our metapopulation approach, these partial differential equation (PDEs) are derived in the continuum limit at the lowest order of a size expansion in $N$ of the Markov chain associated with the processes (\ref{sel})-(\ref{migr})~\cite{TuringNoisy,inprep}.

\begin{figure}
		\includegraphics[width=0.99\linewidth]{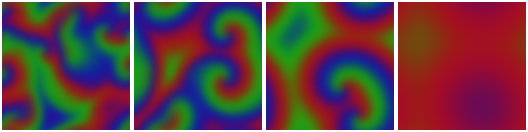}\\
		~\\
		\includegraphics[width=0.95\linewidth]{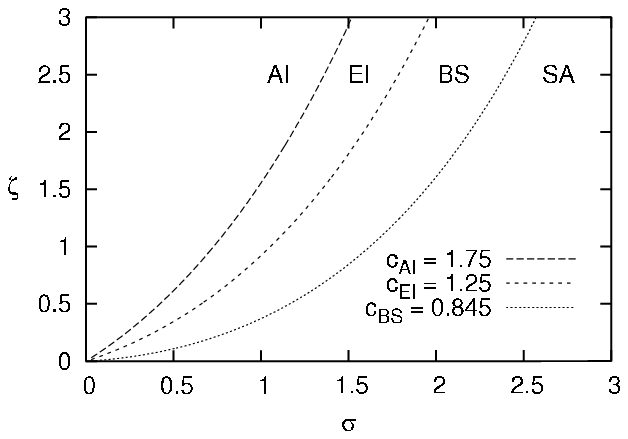}
		\caption{Upper: Typical snapshots from the PDE (\ref{PDE}) in phases AI, EI, BS, SA from left to right (compare with Fig.~\ref{ssa_four_phases}, same parameters used). Lower: System's phase diagram around the HB with contours of $c = (c_{AI}, c_{EI}, c_{BS})$ and $\beta = 1$. As a comprehensive feature, we distinguish four phases: spiral waves are unstable in AI, EI and SA, but are stable in BS (see text).}
		\label{c3_state_diagram}
	\end{figure}

Here, we aim to unravel the combined influence of nonlinearity, mobility and noise on the system's dynamics and the formation and stability of coherent patterns. To gain some insight into these questions, we report some typical lattice simulations (performed using the Gillespie algorithm~\cite{Gillespie,inprep}) obtained in the regime where there is a limit cycle ($\mu < \mu_H$). As shown in Fig.~\ref{ssa_four_phases}, this parameter regime is characterised by spiralling patterns found in four different phases (i.e. four parameter regimes), whereas we have found no patterns when $\mu > \mu_H$ (see~\cite{FigshareMovies}). We have checked that the PDEs (\ref{PDE}) faithfully reproduce the behaviours obtained with Gillespie lattice simulations of the metapopulation model (\ref{sel})-(\ref{migr}) as shown in Fig.~\ref{c3_state_diagram}~(upper) and~\cite{FigshareMovies}.

\section{Asymptotic expansion}
The main goal of this work is to obtain an analytical description of the metapopulation model's phase diagram and an understanding of the circumstances under which the spiralling patterns of Fig.~\ref{ssa_four_phases} are stable or unstable. Our approach relies on the description of the metapopulation system by the PDEs (\ref{PDE}) whose properties near the HB will be studied perturbatively (see below). For this, it is convenient to perform the linear transformation ${\bm s}-{\bm s}^*\to (u,v,w)$, with $u=-(r+s_3)/\sqrt{6}$, $v=(s_2-s_1)/\sqrt{2}$ and $w=r/\sqrt{3}$. In these variables, the linear part of (\ref{RE}) can be written in the Jordan normal form $\partial_t (u + i v) = (\epsilon^2 + i\omega_H)(u + iv)$ and $\partial_t w = -\beta w$.

To make analytical progress and following a classic asymptotic approach, see e.g.~\cite{Manneville,Miller}, we perform a space and time perturbation expansion in the parameter $\epsilon$ around the HB. For this, we introduce the multiple scale coordinates $T = \epsilon^2 t$ and $\bm{X} = \epsilon \bm{x}$ with $\Delta_{\bm X} \equiv \partial^2_{X_1} + \partial^2_{X_2}$, and expand the densities in powers of $\epsilon$. This yields
	\begin{equation}
		\label{perturbation}
		 u(\bm{x},t) = \sum_{n=1}^{3} \epsilon^n U^{(n)}(t,T,\bm{X})
	\end{equation}
and, similarly, $v=\sum_{n=1}^{3} \epsilon^n V^{(n)}$ and $w= \sum_{n=1}^{3} \epsilon^n W^{(n)}$, where the functions $U^{(n)}, V^{(n)}, W^{(n)}$ are of order ${\cal O}(1)$. Substituting (\ref{perturbation}) into (\ref{PDE}) and, using the definition of $(u,v,w)$, we obtain a hierarchy of PDEs and analyse them at each order of $\epsilon$. Since the variables $u$ and $v$ are decoupled from $w$ at linear order, one writes $U^{(1)}+iV^{(1)}=\mathcal{A}(T,\bm{X})e^{i\omega_H t}$, where $\mathcal{A}$ is the complex modulation amplitude. The decoupled equations for $w$ give $W^{(1)}=0$ and $W^{(2)} \propto |\mathcal{A}|^2$, which is the leading term in the equation for the centre manifold~\cite{GuckenheimerHolmes}. To obtain a sensible expansion all secular terms are removed. A first such term arises at order $\mathcal{O}(\epsilon^3)$ and its removal yields the complex Ginzburg--Landau equation (CGLE)~\cite{CGLE} with a real diffusion coefficient $\delta$
	\begin{equation}
	\label{CGLE}
		\partial_{T} \mathcal{A} =
		\delta \Delta_{\bm X} \mathcal{A} + \mathcal{A} - (1 + i c) |\mathcal{A}|^2 \mathcal{A},
	\end{equation}
where $\delta = \frac{3\beta\delta_E + \sigma\delta_D}{3\beta + \sigma}$ and $\mathcal{A}$ has been rescaled by a constant to give 
	\begin{equation}
	\label{c3}
		c = \frac
		{12\zeta (6\beta - \sigma)(\sigma + \zeta) + \sigma^2 (24\beta - \sigma)}
		{3\sqrt{3} \sigma (6\beta + \sigma)(\sigma + 2\zeta)}.
	\end{equation}
We emphasize that the CGLE (\ref{CGLE}) has been derived here in a {\it controlled} perturbative expansion and describes the system's dynamics to order $\epsilon$ near the HB. This treatment, therefore, differs from that of Refs.~\cite{RMF,Matti,RF,Jonas}, where CGLEs were obtained by heuristically treating heteroclinic cycles as limit cycles.

	\begin{figure}
		\onefigure[width=0.75\linewidth]{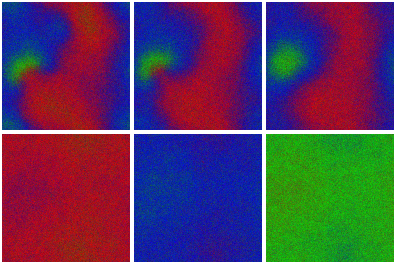}
		\caption{Typical time evolution of the stochastic system in the SA phase. Same parameters and initial conditions as in Fig.~\ref{ssa_four_phases} with $\zeta = 0$. Upper: spiral annihilation at different stages, for time $t=(234,310,386)$ from left to right. Lower: the oscillatory dominance of each species at $t=(955,967,980)$ after relaxation into the homogeneous state (no species extinction).}
		\label{ssa_spiral_annihilation}
	\end{figure}

\section{Phase diagram and CGLE}
According to the CGLE (\ref{CGLE}), the movement in the vicinity of the HB is described by {\it linear} diffusion, with an effective diffusion constant $\delta$ depending on $\delta_D$ and $\delta_E$ (\ref{migr}). When reproduction dominates over selection ($\beta \gg \sigma$), the lack of empty spaces leads to prevalence of pair-exchanges ($\delta \to \delta_E$), while in the opposite case ($\beta \ll \sigma$), movement occurs mostly via hopping ($\delta \to \delta_D$). As the effective linear diffusive term in (\ref{CGLE}) affects only the size of the patterns but not their stability, for our purpose here $\delta$ can be always rescaled to 1 via ${\bm x} \to {\bm x}/\sqrt{\delta}$. In addition, one of the three parameters ($\beta$, $\sigma$, $\zeta$) can always be set to 1 by an appropriate rescaling of time (we have here chosen to set $\beta = 1$), while $\mu \approx \mu_H$ since we consider an expansion near the HB. Therefore, the phase diagram around the HB (represented in Fig.~\ref{c3_state_diagram}) can comprehensively be described in terms of $\sigma$ and $\zeta$ with $\beta = 1$.

The system's phase diagram near the HB (Fig.~\ref{c3_state_diagram}, see also the movies of \cite{FigshareMovies}) is the main result of this work and has been inferred from (\ref{CGLE}) and (\ref{c3}) by referring to the well-known properties of the two-dimensional CGLE~\cite{CGLE}. This phase diagram is characterised by four phases with three critical values of $c$, as illustrated in Fig.~\ref{c3_state_diagram}. In the ``spiral annihilation'' (SA) phase, when $0 < c < c_{BS}$, the dynamics is characterised by unstable spiralling patterns that collide and vanish. In the ``bound state'' (BS) regime $c_{BS} < c < c_{EI}$, pairs of stable spirals are formed and coevolve, with their properties described by the CGLE (\ref{CGLE})~\cite{inprep}: e.g., the speed and wavelength of the spiral waves grow $\propto\sqrt{\delta}$.
When $c_{EI} < c < c_{AI}$, the spirals become convectively unstable due to the Eckhaus instability (EI) which limits their size and distorts their shape. It is noteworthy that EI has been reported in~\cite{RF} for a model without mutations ($\mu = 0$). Finally, there is the ``absolute instability'' (AI) of spiral waves when $c_{AI} < c$, where there are no coherent patterns since the cores are not able to sustain spiral arms. By substituting the explicit values $c_{BS} \approx 0.845$, $c_{EI} \approx 1.25$ and $c_{AI} \approx 1.75$~\cite{CGLE} into (\ref{c3}), one obtains the system's phase diagram in the $\sigma-\zeta$ plane as shown in Fig.~\ref{c3_state_diagram}. This phase diagram sheds light on the results of Fig.~\ref{ssa_four_phases} where the values $\zeta = (1.8, 1.2, 0.6, 0)$ correspond to $c = (1.9, 1.5, 1.0, 0.6)$, which lie in the four phases AI, EI, BS and SA respectively. A description of the evolution in each phase can be found in the accompanying movies~\cite{FigshareMovies}.
The SA phase (see Fig.~\ref{ssa_spiral_annihilation}), which was not found in Refs.~\cite{RMF,RF,Matti,He2,Jonas}, is characterised by the annihilation of all spiralling patterns and is particularly interesting since it is the only possible phase near the HB when $\zeta = 0$ (see Fig.~\ref{c3_state_diagram}), i.e. for the models of~\cite{RMF,He2} supplemented by mutations. In this novel SA phase, spiral annihilation leads to a spatially-homogeneous oscillating state dominated in turn by each species, without any of them going extinct, as described by the mean field dynamics (\ref{RE}). This deterministic phenomenon (different from the EI) is driven by nonlinearity and not by demographic noise. In the regime $c \ll c_{BS}$, it typically occurs on a short time scale, as illustrated in Fig.~\ref{ssa_spiral_annihilation}. This is markedly different from the loss of spiralling patterns driven by noise after a time growing exponentially with the system size as found in~\cite{RMF,He2}.

\begin{figure}
		\onefigure[width=0.99\linewidth]{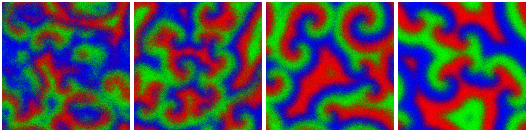}
		\caption{Stochastic simulations with the same initial conditions and same parameters as in Fig.~\ref{ssa_four_phases}, but with a low mutation rate $\mu=0.001$. While the AI, EI and BS are still present in agreement with the phase diagram of Fig.~\ref{c3_state_diagram} (see caption of Fig.~\ref{c3_state_diagram} for the order of the phases), no spiral annihilations occur and the SA phase of Fig.~\ref{ssa_four_phases} is now replaced by the BS phase.}
		\label{ssa_four_phases_low_mutation}
	\end{figure}

While our analysis in terms of the CGLE (\ref{CGLE}) relies on a perturbative treatment around the HB where $\epsilon \ll 1$, it is still found to faithfully describe the system's properties relatively far from the HB. For instance, when $\beta=\sigma=1$ and $\zeta=0$ ($\mu_H=0.042$), the system is still in the SA phase even for $\mu=0.02$ ($\epsilon \approx 0.26$) as predicted by our theory (see Figs.~\ref{ssa_four_phases} and \ref{ssa_spiral_annihilation}). We have also found that the predictions for the existence of the AI, EI and BS phases still hold even for quite low mutation rates, as illustrated in Fig.~\ref{ssa_four_phases_low_mutation}: when $\mu$ is much smaller than the rates $\sigma$ and $\beta$ (e.g. $\mu = 0.001$ and $\sigma=\beta=1$), the system lies in the AI, EI and BS phases as predicted by the phase diagram of Fig.~\ref{c3_state_diagram}. However, no spiral annihilation occurs in such a regime ($\epsilon\approx 0.35$) and instead one finds stable spiralling patterns (rightmost panel of Fig.~\ref{ssa_four_phases_low_mutation}). In agreement with the phase diagram of Fig.~\ref{c3_state_diagram}, the system is in the AI phase when $0 < \sigma \ll \zeta$ (leftmost panel of Fig.~\ref{ssa_four_phases_low_mutation}), including when $\sigma$ and $\mu$ are small and $\zeta$ is finite. It is interesting to note that no stable spiralling patterns have been found in a two-dimensional zero-sum variant of the model, with $N=1, \sigma=\mu=\beta=0$ and $\zeta=1$~\cite{Matti}.

 We have also checked that our analysis is robust against simultaneous random perturbations (up to $\pm5\%$) of all the reaction rates (\ref{sel})-(\ref{migr})~\cite{inprep}. As shown in Figs.~\ref{ssa_four_phases} and \ref{c3_state_diagram}, the PDEs (\ref{PDE}) describe perfectly the stochastic metapopulation model when $N \gg 1$ and, in practice, are still accurate when $N \gtrsim 16$ for any nonzero mobility. Furthermore, when $N = 2$ and the mobility rates are sufficiently high~\cite{RMF}, the phase diagram of Fig.~\ref{c3_state_diagram} is still valid~\cite{FigshareMovies}.

\section{Nonlinear mobility}
\begin{figure}
		\onefigure[width=0.750\linewidth]{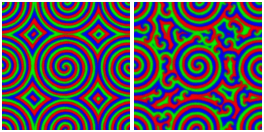}
		\caption{Influence of mobility on spiralling patterns: typical snapshots in Gillespie lattice simulations for $(\delta_D,\delta_E)=(0.05,0.05), (0.20,0.05)$ from left to right respectively. Other parameters are: $L^2=128^2$, $N=1024$, $\beta=\sigma=1$, $\zeta=0.1$ and $\mu=10^{-6} \ll \mu_H=0.042$. Geometrically ordered initial conditions, see movies~\cite{FigshareMovies} for full details.}
		\label{ssa_different_mobilities}
	\end{figure}

Near the HB, we have seen that the dynamics is aptly captured by the CGLE~(\ref{CGLE}) with linear diffusion and the system's properties can be described without loss of generality by setting $\delta_D=\delta_E$ (see Figs~\ref{ssa_four_phases} and \ref{c3_state_diagram}). This is no longer the case when the mutation rate is very low (e.g. for $\mu = 10^{-6}$ as in Fig.~\ref{ssa_different_mobilities}) and the dynamics is then far away from the HB. In fact, 
the CGLE (\ref{CGLE}) does no longer provide a quantitatively detailed description of the dynamics
in the regime of very low mutation rate, where the SA phase is replaced by a phase characterised by spiralling patterns whose stability is affected by the nonlinear diffusive terms of (\ref{PDE}). As an illustration, in Fig.~\ref{ssa_different_mobilities} we show that a far-field break-up of the spiralling patterns solely caused by nonlinear mobility occurs when $\delta_D \ne \delta_E$ and $\mu\ll \mu_H$ (the coexistence state ${\bm s}^*$ is unstable) and the noise intensity is negligible (since $N \gg 1$), see also~\cite{FigshareMovies}.

\section{Conclusion}
In summary, we have investigated the stability of spiralling patterns in a generic three-species model whose evolution results from the combined biologically-motivated effects of cyclic dominance, mutation and nonlinear mobility. Inspired by recent experiments~\cite{Nahum}, we have developed a metapopulation description and analysed the dynamics in terms of PDEs and the CGLE derived from a size expansion and a multiscale perturbative treatment around the Hopf bifurcation, and by simulations with stochastic Gillespie algorithm. We have thus obtained the system's phase diagram, which is characterised by four phases, with only one capable of supporting stable spiralling patterns. The instabilities in the three other phases are not driven by noise. In particular, we have identified a phase (SA) where spirals annihilate, leading to spatially uniform dominance of each species in turn. Importantly, these behaviours, which arise in a wide region of the parameter space around the Hopf bifurcation, are robust and {\it independent} of the mobility rates. This is in stark contrast with the results of Refs.~\cite{RMF,RF,He2}, where spiralling patterns and spatial uniformity were respectively found at low and high mobility, and may explain why spiralling patterns turn out to be elusive in the microbial experiments of Refs.~\cite{Kerr,Nahum}. We have also shown that, regardless of internal noise and beyond the range of validity of the CGLE, nonlinear diffusion causes far-field break-up of spiral waves away from the Hopf bifurcation when the coexistence state is unstable.

While we have here focused on a two-dimensional (square lattice) metapopulation model, which is a setting particularly relevant to model the co-evolution of microbial communities~\cite{Kerr,Nahum}, it is worth noting that the dynamics of models closely related to the RPS games have also been studied on random and complex networks, see e.g. \cite{OnNetworks}. It would therefore be interesting, for instance, to investigate whether our theoretical approach can help shed further light on the properties of the oscillating patterns characterising some RPS games on small-world networks~\cite{RPS-SW}.

\acknowledgments
The authors acknowledge discussions with Tobias Galla at the early stage of this project. BS is grateful for the support of an EPSRC studentship.


\begin{thebibliography}{99}
\bibitem{spatio-temporal}
	Murray, J. D., {\it Mathematical Biology} (Springer-Verlag, New York, 1993);
	Koch, A. J., \and Meinhardt, H., Rev. Mod. Phys. {\bf 66}, 1481 (1994).
\bibitem{patterns}
	Levin, S. A. \and Segel, L. A., Nature (London) {\bf 259}, 659 (1976);
	Hassel, M. P., Comins, H. N. \and May, R. M., {\it ibid} {\bf 370}, 290 (1994);
	Abraham, E. R., {\it ibid} {\bf 391}, 577 (1998);
	Maron, J. L. \and Harrison, S., Science {\bf 278}, 1619 (1997).
\bibitem{PatternsRPS}
	Jackson, J. B. C. \and Buss, L., Proc. Natl. Acad. Sci. U.S.A. {\bf 72}, 5160 (1975);
	Sinervo, B. \and Lively, C. M., Nature (London) {\bf 380}, 240 (1996);
	Kirkup, B. C. \and Riley, M. A., {\it ibid.} 428, 412 (2004).
\bibitem{Turing}
	Turing, A.  M., Phil. Trans. R. Soc. B {\bf 237}, 37 (1952).
\bibitem{TuringNoisy}
	Lugo, C. A. \and McKane, A. J., Phys. Rev. E {\bf 78}, 051911 (2008);
	Butler, T. \and Goldenfeld, N., Phys. Rev. E {\bf 80}, 030902(R) (2009); {\it ibid.} {\bf 84}, 011112 (2011).
\bibitem{biodiversity}
	Pennisi, E., Science {\bf 309}, 93 (2005);
	Thuiller, W., Nature 448, 550 (2007).
\bibitem{Kerr}
	Kerr, B., Riley, M. A., Feldman, M. W. \and Bohannan, B. J. M., Nature (London) {\bf 418}, 171 (2002).
\bibitem{Nahum}
	Kerr, B., Neuhauser, C., Bohannan, B. J. M. \and Dean, A. M., Nature {\bf 442}, 75 (2006); 
	Nahum, J. R., Harding, B. N. \and Kerr, B., Proc. Natl. Acad. Sci. U.S.A. {\bf 108}, 10831 (2011).
\bibitem{LizardMutation}
	Sinervo, B., Miles, D. B., Frankino, W. A., Klukowski, M. \and DeNardo, D. F., Hormones and Behaviour {\bf 38}, 222 (2000).
\bibitem{games}
	Hofbauer, J. \and Sigmund, K., {\it Evolutionary games and population dynamics} (Cambridge University Press, 1998);
	Frey, E., Physica A {\bf 389}, 4265 (2010).
\bibitem{RPSnonspatial}
	Reichenbach, T., Mobilia, M. \and Frey, E., Phys. Rev. E {\bf 74}, 051907 (2006);
	Berr, M.,Reichenbach, T., Schottenloher, M. \and Frey, E., Phys. Rev. Lett. {\bf 102}, 048102 (2009);
	Mobilia, M., J. Theor. Biol. {\bf 264}, 1 (2010);
	M\"uller, A. P. O. \and Gallas, J. A. C., Phys. Rev. E {\bf 82}, 052901 (2010).
\bibitem{zerosum}
	Tainaka, K. I., Phys. Rev. Lett. {\bf 63}, 2688 (1989); Phys. Rev. E {\bf 50}, 3401 (1994);
	Frachebourg, L., Krapivsky, P. L. \and Ben-Naim, E., Phys. Rev. Lett. {\bf 77}, 2125 (1996);
	Szab\'o, G., Szolnoki, A. \and Izs\'ak, R., J. Phys. A. Math. Gen. {\bf 37}, 2599 (2004);
	Szab\'o, G. \and Szolnoki, A., Phys. Rev. E {\bf 65}, 036115 (2002);
	He, Q., Mobilia, M. \and T\"auber, U. C., {\it ibid.} {\bf 82}, 051909 (2010);
	Perc, M., Szolnoki, A. \and Szab\'o, G., {\it ibid.} {\bf 75}, 052102 (2007);
	Ni, X., Wang, W. X., Lai, Y. C. \and Grebogi, C., {\it ibid.} {\bf 82}, 066211 (2010);
	Jiang, L., Zhou, T., Perc, M., Huang, X. \and Wang B., New J. Phys. {\bf 11}, 103001 (2009).
\bibitem{RMF}
	Reichenbach, T., Mobilia, M. \and Frey, E., Nature (London) {\bf 448}, 1046 (2007); Phys. Rev. Lett. {\bf 99}, 238105 (2007); J. Theor. Biol. {\bf 254}, 368 (2008).
\bibitem{RF}
 	Reichenbach, T. \and Frey, E., Phys. Rev. Lett. {\bf 101}, 058102 (2008).
\bibitem{Matti}
	Peltom\"aki, M. \and Alava, M., Phys. Rev. E {\bf 78}, 031906 (2008). 
\bibitem{He2}
	He, Q., Mobilia, M. \and T\"auber, U. C., Eur. Phys. J. B {\bf 82}, 97 (2011);
	He, Q., T\"auber, U. C. \and Zia, R. K. P., {\it ibid.} {\bf 85}, 141 (2012).
\bibitem{MayLeonard}
	May, R. M. \and Leonard, W. J., SIAM J. Appl. Math. {\bf 29}, 243 (1975). 
\bibitem{Jonas}
	Cremer, J., MSc Thesis (Ludwig-Maximilians-Universit\"at M\"unchen, 2007).
\bibitem{metapopulation}
	Levins, R., Bull. Entomol. Soc. Am. 15, 237 (1969);
	Hanski, I., {\it Metapopulation Ecology} (New York, Oxford University Press, 1999).
\bibitem{FigshareMovies}
	Szczesny, B., Mobilia, M. \and Rucklidge, A. M., figshare, doi: \href{http://dx.doi.org/10.6084/m9.figshare.96949}{10.6084/m9.figshare.96949}
\bibitem{Kearns}
	Kearns, D. B., Nature Rev. Micro. {\bf 8}, 634 (2010).
\bibitem{Gillespie}
	Gillespie, D.~T., J. Comput. Phys. {\bf 22}, 403 (1976).
\bibitem{inprep}
	Szczesny, B., Mobilia, M. \and Rucklidge, A. M., in preparation.
\bibitem{SizeExp}
	Van Kampen, N. G., {\it Stochastic Processes in Physics and Chemistry} (Elsevier, 2007);
	Gardiner, C., {\it Stochastic Methods} (Springer, 2010).
\bibitem{Fanelli}
	Fanelli, D., Cianci, C. \and Di Patti, F., arXiv: \href{http://arxiv.org/abs/1112.0870v2}{1112.0870v2}
\bibitem{Manneville}
	Manneville, P., {\it Dissipative structures and weak turbulence} (Academic Press, San Diego, 1992).
\bibitem{Miller}
	Miller, P., {\it Applied Asymptotic Analysis, Graduate Studies in Mathematics} (American Mathematical Society, 2006).
\bibitem{GuckenheimerHolmes}
	Guckenheimer, J. \and Holmes, P., {\it Nonlinear Oscillations, Dynamical Systems, and Bifurcations of Vector Fields} (Springer, 1983).
\bibitem{CGLE}
	Aranson, I. S., Kramer, L. \and Weber, A., Phys. Rev. E {\bf 47}, 3231 (1993);
	Aranson, I. S. \and Kramer, L., Rev. Mod. Phys. {\bf 74}, 99-143 (2002).
\bibitem{OnNetworks}
	Szab\'o, G., F\'ath, G., Phys. Rep. {\bf 446}, 97 (2007);
	Perc, M. \and Szolnoki, A., BioSystems {\bf 99}, 109-125 (2010).
\bibitem{RPS-SW}
	Szab\'o, G., Szolnoki, A. \and Izs\'ak, R., J. Phys. A: Math. Gen. {\bf 37}, 2599 (2004).
 
\end{thebibliography}
\end{document}